\documentclass[review]{elsarticle}

\usepackage{siunitx}
\usepackage{float}
\usepackage{amsmath}
\usepackage{amsthm}
\usepackage{caption}
\usepackage{subcaption}
\usepackage[utf8]{inputenc} % allow utf-8 input
\usepackage[T1]{fontenc}    % use 8-bit T1 fonts
\usepackage{hyperref}       % hyperlinks
\usepackage{url}            % simple URL typesetting
\usepackage{booktabs}       % professional-quality tables
\usepackage{amsfonts}       % blackboard math symbols
\usepackage{nicefrac}       % compact symbols for 1/2, etc.
\usepackage{microtype}      % microtypography
\usepackage{lipsum}
\usepackage{graphicx}

\begin{document}
\title{Thermal spin injection from a ferromagnet into graphene by transverse and longitudinal current}

\author{
 Bin Yang \\  
 Winfried Teizer \\
  Department of Physics\&Astronomy \\
  Texas A\&M University \\
  College Station, TX 77840 }
%   \texttt{abel9394@tamu.edu} \\
  %% examples of more authors

%   \texttt{teizer@tamu.edu} \\

% \title{This is some thing}

\begin{abstract}
Graphene is a very promising material in spintronics due to both its high electric mobility and low intrinsic spin-obit coupling. Electronic spins can be injected from a ferromagnetic material through a tunnel contact into graphene owing to a spin relaxation length as high as 5$\mu m$. In recent years, a new approach creating spin current employed thermal effects and heat flow. Here, by applying transverse and longitudinal current to a grahene spin valve device,  the interplay between the heat spin current and the charge spin current is investigated. The non-local spin voltage is enhanced by the thermal spin injection and  the thermal spin voltage reaches a maximum close to the Dirac point which makes graphene a promising material for a future thermoelectric spin device due to its long spin lifetime and spin diffusion length.   

\end{abstract}
\maketitle

\section{Background}
Spintronics, short for spin electronics, is the study of coupled electron spin and electronic charge in condensed matter physics. The application of spintronics in magnetic hard drives and other storage media based on GMR and TMR \cite{gmr} has achieved great success. A topic in spintronics is the injection of spin polarized electrons from ferromagnetic materials into non-magnetic materials, such as semiconductors \cite{spininjection}. This raises the possibility to create various spintronics devices with low energy consumption, based on spin currents. Besides spin transport, heat current is also playing an important role in condensed-matter structures and devices. Reducing waste energy is a challenge for making efficient electronic devices. The generated heat in the transistors is also one of the main factors causing the breakdown of Moore's law. Spintronics is one of the methods to reduce the energy consumption while the thermoelectric effect may help to convert heat into electric power and enhance spin transport. The interplay between heat and spins attracted increased interest to the so-called spin caloritronics \cite{spincaloritronics}.

The interplay between heat and charge currents in metallic materials is described as the Seebeck effect \cite{seebeck}\cite{seebeck2}, which was first discovered in 1794 by Italian scientist Alessandro Volta, and German physicist Thomas Johann Seebeck independently rediscovered it in 1821. It states that if there is a thermal gradient applied between two ends of a metal, the electrons will flow from the hot end to the cold end and an electric field builds up, which can be written as $\vec{E}=-S\nabla T$. $S$ is the Seebeck coefficient and can be linearly approximated by the Sommerfeld expansion, which is called the Mott relation. If we consider both the electric and thermal gradient in the metal, the electric current can be expressed as $\vec{j_{c}}=-\sigma\nabla V-\sigma S\nabla T$. The heat current can be expressed as $j_{q}=-\Pi\sigma\nabla V-\kappa\nabla T$, where $\Pi$ is the Peltie coefficient and $\kappa$ is the thermal conductivity. To generalize the spin current, the two-spin current model is applied. Charge current and spin current is expressed as $j_{c}=j^{\uparrow}+j^{\downarrow}$ and $j_{s}=j^{\uparrow}-j^{\downarrow} $ respectively. If we define the spin chemical potential $\mu_{s}=\mu^{\uparrow}-\mu^{\downarrow}$, spin current can be written as $j_{s}=-\frac{\sigma}{e}\nabla\mu_{s}$. In general, different types of current can be written in matrix form \cite{spincaloritronics}:
\begin{equation}
\begin{pmatrix}
j_{c}\\
j_{s}\\
j_{q}
\end{pmatrix} =\sigma
\begin{pmatrix}
1 & P & ST\\
P & 1 & P^{'}ST\\
ST & P^{'}ST & \kappa T/\sigma 
\end{pmatrix}
\begin{pmatrix}
\nabla\mu_{c}/e\\
\nabla\mu_{s}/2e\\
-\nabla T/T

\end{pmatrix}
\end{equation}
where $P=\frac{\sigma^{\uparrow}-\sigma^{\downarrow}}{\sigma}|_{\varepsilon=\varepsilon_{F}}$ and $P^{'}=\frac{\partial(P\sigma)}{\partial\varepsilon}|_{\varepsilon=\varepsilon_{F}}$.
Our experiment is mainly related to the spin-dependent Seebeck effects \cite{spinseebeck}. In FM|NM metallic spin valve structures, spin injection from a ferromagnet into a non-magnetic material generates different carrier densities between spin up and spin down electrons $n=n^{\uparrow}-n^{\downarrow}$ according to the two-spin current model, and results in the spin accumulation $\mu = \mu^{\uparrow}-\mu^{\downarrow}$ at the interface. The Seebeck coefficient is different in case of two spin sub-bands due to different carrier densities $S^{\uparrow}\neq S^{\downarrow}$, which is called the spin Seebeck effects. If we apply both the electric and thermal gradient on the spin-valve structure, the change of the chemical potential consists of two parts: the first one is caused by the electric spin potential $\delta\mu_{c}$, which is related to the spin diffusion length and spin accumulation on the interface, while another is caused by the thermal spin potential $\delta\mu_{th}=-\Delta S
\nabla T$, which is related to the spin Seebeck effect and can be controlled by the carrier density. 

In recent years, spin  caloritronics attracts more and more interest, including in the injection of thermal spin current from ferromagnets into semiconductors \cite{thermalsemi}\cite{thermalsemi2}. With the help of thermal spin currents, the spin accumulation at the interface may be enhanced, which indicates that dissipated heat can be used as a boost in spintronics. Here, we demonstrate that graphene, like semiconductors, can be injected with both thermal spins and electric spins \cite{graphenespin} by both transverse and longitudinal currents. With the gate voltage applied (adjust the Fermi energy), the total spin accumulation can be enhanced by thermal spin injection and reaches a maximum close to the Dirac point.

\section{Graphene spin valve device}
\begin{figure}

 \includegraphics[width=\columnwidth]{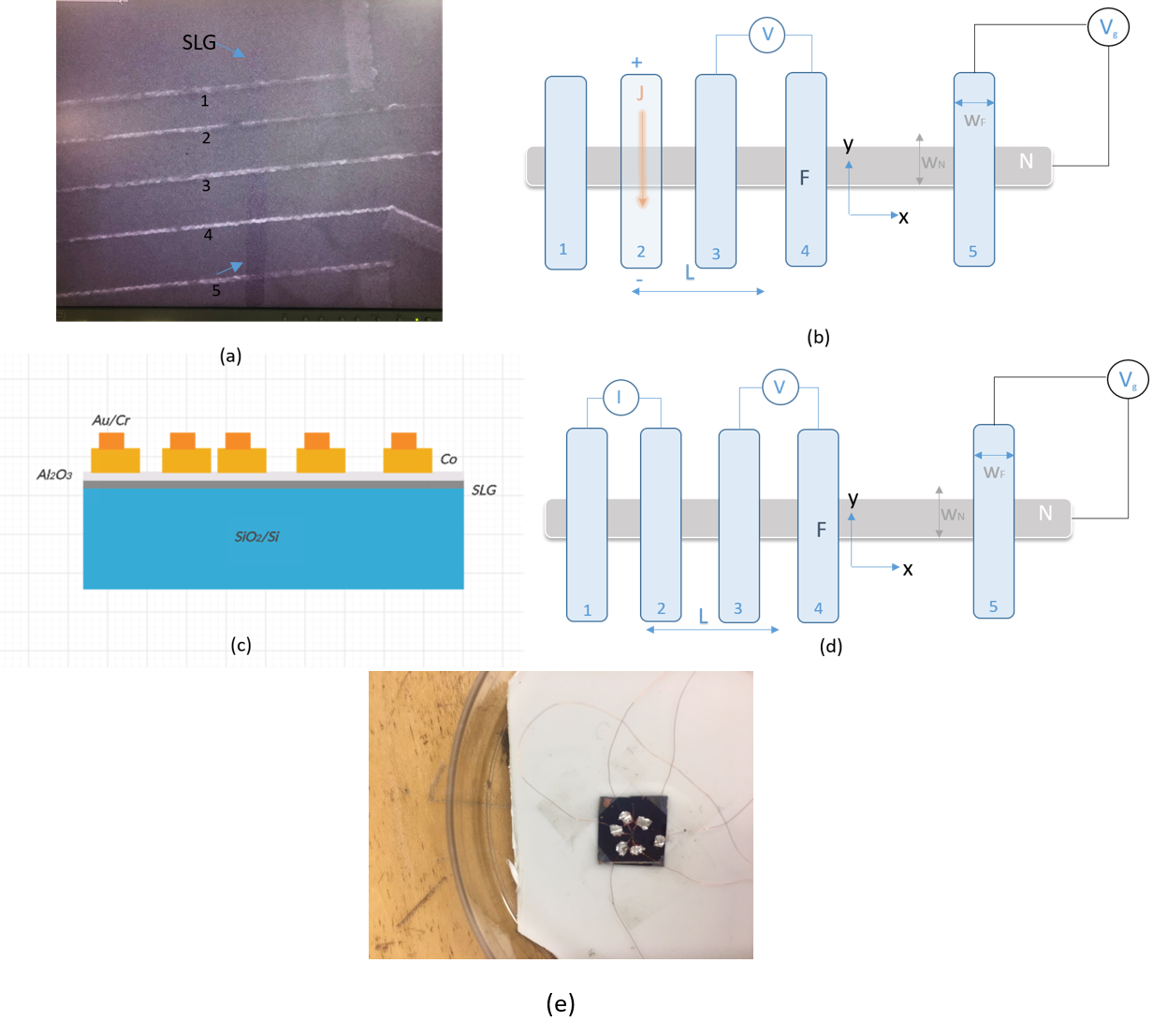}

\caption{(a) SEM image of spin valve device. (b) Transverse current applied through contact 2 on the spin valve device. (c) Spin valve device in the vertical view.  (d) Longitudinal current applied from contact 2 to contact 1 on the spin valve device. (e) Graphene spin valve device sample. } 
\label{fig:config}
\end{figure}

 Graphene spin valve devices are fabricated in three steps as shown in Fig.\ref{fig:config}(c): First a single layer graphene (SLG) flake is fabricated from ZYA grade highly oriented pyrolytic graphite (HOPG) by mechanical exfoliation and identified by optical microscopy. Raman spectroscopy is utilized to further determine the number of layers and finally locate the position of SLG. Then electron beam lithography is used to deposit a $1nm$ ultra-thin $Al_{2}O_{3}$ tunnel barrier on top of the graphene flake by magnetron sputtering. At last, ferrromagnetic Cobalt contacts are deposited on top of the tunnel barrier by thermal evaporation. The SEM image of our samples are shown in Fig.\ref{fig:config}(a). In the experiment, we apply transverse and longitudinal current on the sample to investigate the interplay between charge spin and thermal spin injection as shown in Fig.\ref{fig:config}(b),(d). The transverse current is directly applied on contact 2 to generate the thermal heat and the longitudinal current is applied between contacts 1 and 2 to generate charge current if the current is small. The fabrication details are shown in the Supplement section. 
 
\section{Measurement and Results}
In this section, we will present spin signal data in the sample shown in Fig.\ref{fig:config}. The device consists of five different cobalt strips, connected to gold contacts at the end, on top of a tunnel barrier. The cobalt strips are connected to gold leads. The widths of the four cobalt contacts for spin injection (1-4) are 200nm, 110nm, 130nm and 220nm respectively and the distance between the contacts is 1.2$\mu m$, 1.5$\mu m$, 1.8$\mu m$ and 1.5$\mu m$ respectively. Contact 5 is for the gate voltage adjustment. 

\begin{figure}

 \includegraphics[width=\columnwidth]{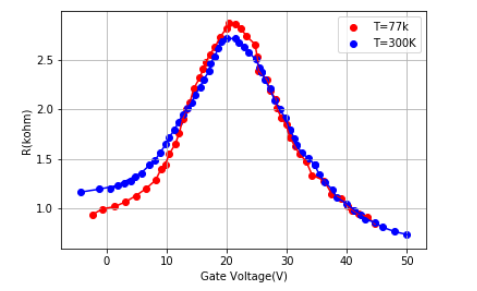}
 \caption{Graphene resistance applied by gate voltage at room temperature and T=77K}
 \label{fig:resist}
\end{figure}

\begin{figure}

 \includegraphics[width=\columnwidth]{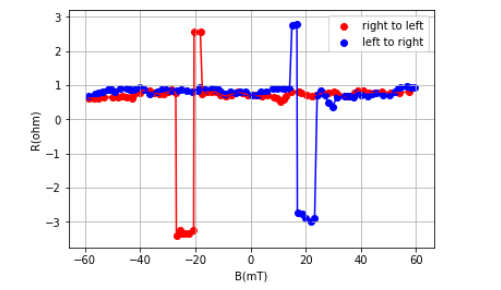}
 \caption{Non-local spin voltage at different magnetic field sweep direction at 77K. The spin direction of four cobalt contacts is $\uparrow\uparrow\uparrow\uparrow\Rightarrow\downarrow\uparrow\uparrow\downarrow\Rightarrow\downarrow\uparrow\downarrow\downarrow\Rightarrow\downarrow\downarrow\downarrow\downarrow$ when magnetic field sweeps from right to left, and $\downarrow\downarrow\downarrow\downarrow\Rightarrow\uparrow\downarrow\downarrow\uparrow\Rightarrow\uparrow\downarrow\uparrow\uparrow\Rightarrow\uparrow\uparrow\uparrow\uparrow$ when magnetic field sweeps from left to right}
 \label{fig:nonlocal}
\end{figure}

The resistivity of graphene, controlled by the gate voltage at room temperature and 77K, is measured by a four-point measurement respectively as shown in Fig. \ref{fig:resist}, where current is injected between contacts 1 and 5 and the voltage is measured between contacts 2 and 4 by a signal recovery 7265 digital lock-in amplifier.  The weak temperature dependence of resistivity is mainly due to an extraordinarily high mobility of charge carriers. To first study the charge spin transport and spin diffusion, a non-local spin voltage measurement is performed. The current is injected from cobalt contact 4 to contact 3 into SLG through the tunnel barrier. With spin-polarized carrier diffusing through SLG, a non-local spin voltage can be detected between contacts 1 and 2 in the ultra-high vacuum chamber, as shown in Fig.\ref{fig:nonlocal} at T=77K. In the Figure, there are two sweep directions of magnetic field. One is from right to left and the positive non-local resistance $R_{NL}=V_{34}/I_{12}$ is due to the flip of spins on contacts 1 and 4, while the negative resistance is owing to the flip of spins on contacts 3. The magnetic non-local resistance $\triangle R_{NL}=R_{\uparrow\uparrow}-R_{\uparrow\downarrow}$ is approximately $4\Omega$. The I-V curves of non-local spin signals at two different temperatures are also measured, as shown in Fig. \ref{fig:IV}. It shows that the current and non-local voltage has a linear relationship at low current bias.

\begin{figure}

\begin{subfigure}{.5\textwidth}
  \centering
  % include first image
  \includegraphics[width=\linewidth]{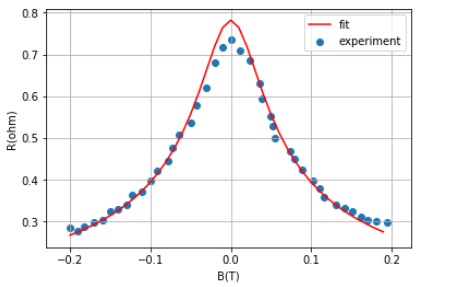}  
  \caption{}
\end{subfigure}
\begin{subfigure}{.5\textwidth}
  \centering
  % include second image
  \includegraphics[width=\linewidth,height=3cm]{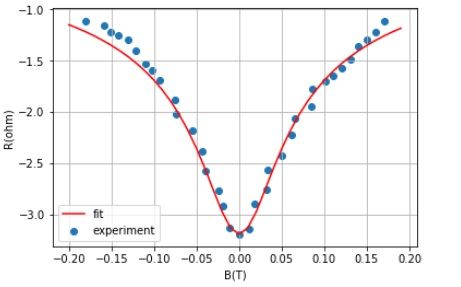}  
  \caption{}
  
\end{subfigure}

\caption{(a) Hanle-type precession measurement in parallel configuration at 77K and fitted to theoretic model (b) Hanle-type precession measurement in anti-parallel configuration at 77K and fitted to theoretic model}
\label{fig:hanle}
\end{figure}

\begin{figure}

 \includegraphics[width=\columnwidth]{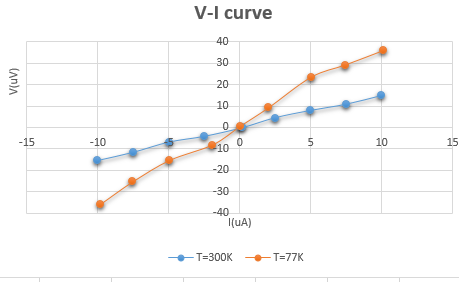}
 \caption{Spin voltage fitting in the linear relationship with respect to different electric current at room temperature and 77K}
 \label{fig:IV}
\end{figure}

To derive the spin relaxation time, spin diffusion length and spin diffusion coefficient, a Hanle precession measurement is also performed. When we apply an out-of-plane magnetic field perpendicular to the magnetization of cobalt, the spin of carriers precesses around the magnetic field during the diffusion process with a Lamor frequency $\Omega=\gamma B$. By assuming that each electron undergoes a random walk from the injector to the destination, we can derive the distribution of electrons on the contacts and the non-local voltage \cite{hanle} $V_{NL}=\pm\int_{0}^{\infty}\frac{1}{\sqrt{4\pi Dt}}\exp({-\frac{L^{2}}{4Dt}})\cos({\Omega t})\exp({-t/\tau_{s}})dt$, where positive implies a parallel and negative an anti-parallel configuration. By fitting the data (Fig. \ref{fig:hanle}) to the model, we find a spin diffusion coefficient $D=1.2\times10^{-2}m^{2}s^{-1}$, spin relaxation time $\tau_{s}=140ps$ and spin diffusion length $\lambda_{s}=\sqrt{D\tau_{s}}=1.3\mu m$ at T=77K. By comparing the non-local resistance with $R_{non-local}=\frac{P^{2}\lambda_{s}}{2W\sigma}\exp({-\frac{L}{\lambda_{s}}})$, we obtain a carrier polarization of about 18\%, which is consistent with other results in the literature \cite{graphenespin}.

To investigate the local thermal spin injection and spin-dependent Seebeck effect, a transverse thermal bias has been established along the cobalt contact by applying a direct current along contact 2. A temperature gradient is evident when the current is raised to 1mA. A thermal spin chemical potential starts to accumulate at the interface due to Seebeck spin tunneling. The general expression for spin accumulation, due to both electric and thermal injection, is as follows \cite{tunnelequation}:
\begin{equation}
    \triangle \mu=\frac{2r_{s}}{R_{tun}+(1-P^{2})r_{s}}[PR_{tun}I-(P_{L}-P_{G})S_{0}\triangle T]
\end{equation}
where $r_{s}$ is the spin resistance of graphene and $R_{tun}$ is the resistance of the tunnel barrier. If we do not apply an electric current by setting I=0A, we obtain a spin potential:
\begin{equation}
    \triangle \mu=\frac{(1-P^{2})r_{s}}{R_{tun}+(1-P^{2})r_{s}}[(S^{\uparrow}-S^{\downarrow})\triangle T]
\end{equation}

\begin{figure}

 \includegraphics[width=\columnwidth]{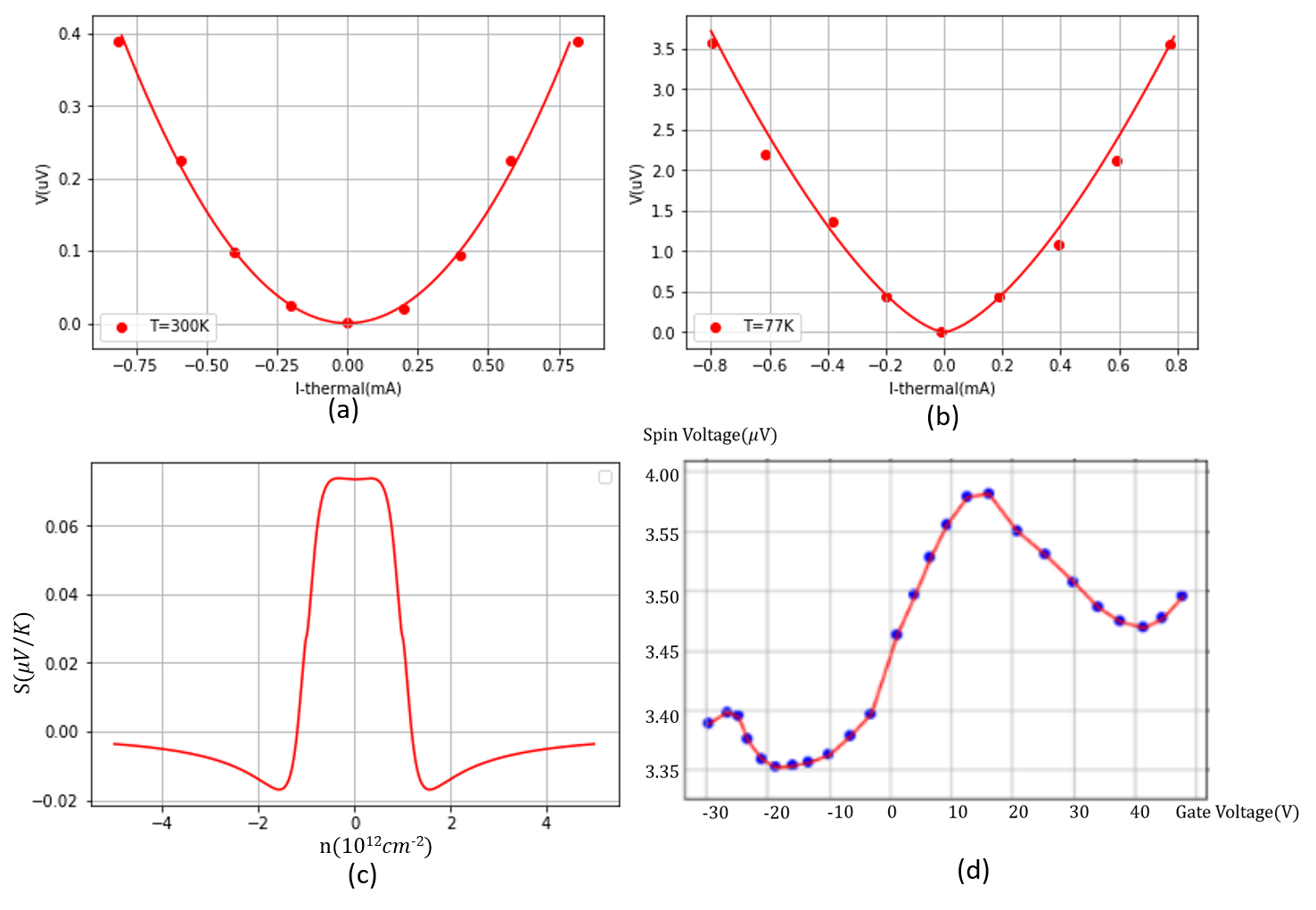}
 \caption{(a) Transverse V-I curve measured at room temperature. (b) Transverse V-I curve measured at T=77K. (c) Theoretical Mott model of derivative of S respect to carrier density. (d) Thermal spin voltage measured with change of gate voltage.}
 \label{fig:transverse}
\end{figure}

The non-local voltage $\delta\mu$, which is defined as the voltage difference between parallel and anti-parallel configuration $V_{non\_local}=V^{\uparrow\uparrow}-V^{\uparrow\downarrow}$, can be obtained from the spin potential due to thermal spin accumulation and a temperature difference between two ends of the graphene flake as:
\begin{equation}
    \delta\mu = -e\frac{\partial(S\Delta T)}{\partial n}\Delta n \propto \frac{\partial S}{\partial n} 
\end{equation}
where $S$ is the Seebeck coefficient of graphene and $\Delta n $ is the carrier density difference due to the spin accumulation at the interface. Here we model our non-local voltage in the simple combination of linear and polynomial form as:
\begin{equation}
    V_{nl} = \alpha I^{\lambda}+\beta I+V_{0}
\end{equation}
where $\alpha$ corresponds to the thermal effect due to the relation between current and temperature, $\beta$ corresponds to the non-local resistivity due to charge current spin injection and $V_{0}$ is the voltage offset due to fluctuations which we ignore here. In the transverse thermal spin injection case, there is no charge spin current injected into graphene. Therefore, the non-local voltage only depends on the first term and $\beta$ can be manually cancelled out. As Fig. \ref{fig:transverse}(a) shows, the non-local spin voltage varies as parabolically with the heat current ($\lambda=2$) at room temperature, as expected from $V_{nl}=\alpha I^{2}$, which is consistent with the traditional relation between current and temperature since the hot carriers in graphene efficiently thermalize with the lattice through scattering processes that they already equilibrate with photons before they reach the detector at room temperature. However, the situation is slightly different at low temperature (T=77K) as shown in Fig. \ref{fig:transverse}(b). The non-local spin voltage is closer to a linear relationship with current ($\lambda$=1.5) since the hot carriers are now not in equilibrium with the photons before they reach the detector at low temperature, which leads to the breakdown of Joule heating relationship. We can also estimate the temperature difference due to the current if we assume a cobalt Seebeck coefficient $S_{c}=-30.8\mu V/K,  \frac{\partial S\Delta T}{\partial n}=8\times 10^{-10}\mu K/ cm^{2} $ and $p=0.3$. The temperature difference between the ferromagnet (Co) and graphene lattice is $\Delta T\approx0.6 K$ at room temperature and $\Delta T\approx 10 K$ at $T=77K$. In addition, we also measure the gate voltage (carrier density) dependence of the non-local voltage as shown in Fig.\ref{fig:transverse}(d). We observe that the trending of non-local spin voltage with respect to carrier density is similar to the trending of $\frac{\partial S}{\partial n}$ as shown in Fig. \ref{fig:transverse}(c) if we assume a Seebeck coefficient S of graphene obeys the Mott relation:
\begin{equation}
    S_{Mott} =\frac{\pi^{2}k_{B}^{2}T}{3e}\frac{\partial lnR}{\partial E},
    R = \mu e\sqrt{n_{0}^{2}+n^{2}}
\end{equation}
where $\mu$ is the carrier mobility in graphene and $n_{0}\approx 3\times 10^{11}cm^{-2}$. We assume it is due to the temperature difference between two ends of graphene sheets which is described in equation (4).

\begin{figure}
 \includegraphics[width=\columnwidth]{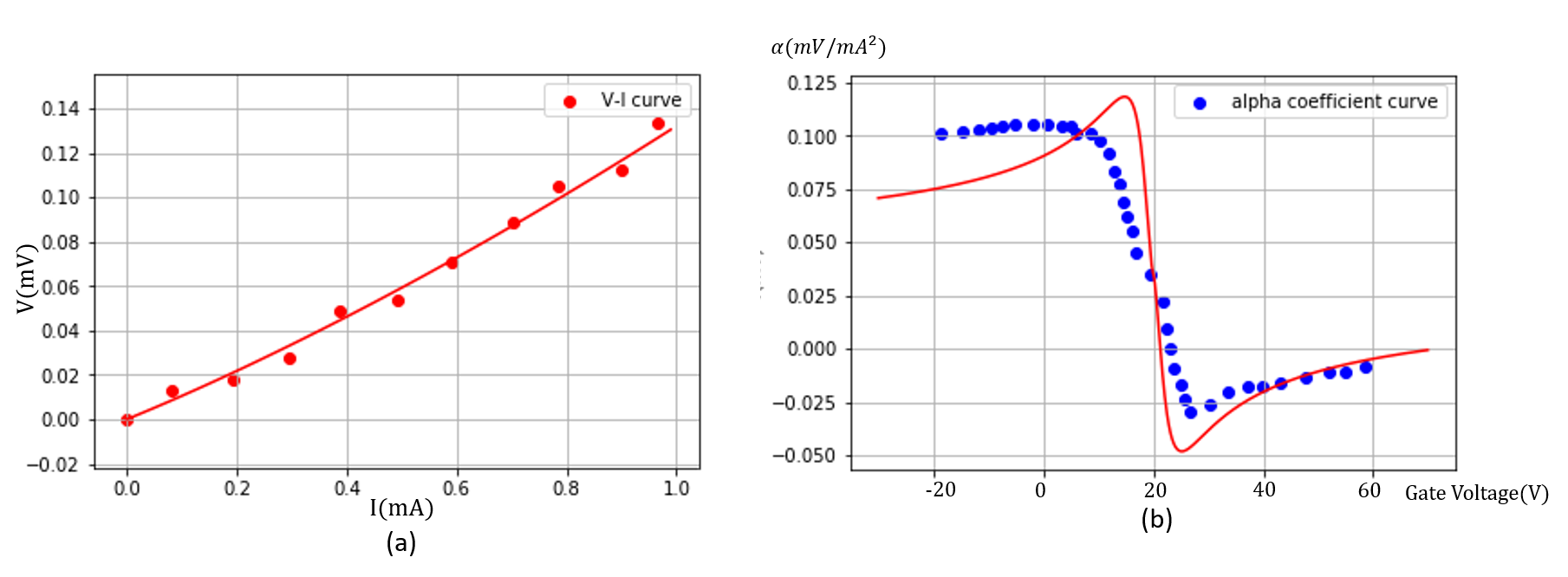}
 \caption{(a) Longitudinal V-I curve measured at T=77K. (b) High order coefficient measured with change of gate voltage and the fitting to the Mott model. }
 \label{fig:longitude}
\end{figure}

In order to estimate the relation between Seebeck coefficient of graphene and carrier density, we also apply a longitudinal current (from contact 2 to contact 1) from the ferromagnetic contact into graphene as shown in Fig.\ref{fig:config}(d). When we gradually increase the longitudinal current up to 1mA, the spin accumulation across the interface contains not only the charge spins but also thermal spins. In this case, we can model our spin voltage as spin up and spin down, which is corresponding to the parallel and anti-parallel configuration:
\begin{gather}
    V_{nl}^{\uparrow}=\alpha^{\uparrow} I^{\lambda}+R^{\uparrow\uparrow}I\\
    V_{nl}^{\downarrow}=\alpha^{\downarrow} I^{\lambda}+R^{\uparrow\downarrow}I
\end{gather}
In which $\alpha^{\uparrow},\alpha^{\downarrow}\propto S^{\uparrow}, S^{\downarrow}$. Since a large longitudinal current causes both charge and thermal spin accumulation, we should keep both the linear and polynomial dependencies of the current with respect to non-local spin voltage and the Seebeck coefficient of graphene can be extract from average of the polynomial coefficients $\Bar{V_{nl}}=\frac{1}{2}( V_{nl}^{\uparrow}+V_{nl}^{\downarrow})=\frac{\alpha^{\uparrow}+\alpha^{\downarrow}}{2}I^{\lambda}+\frac{R^{\uparrow\uparrow}+R^{\downarrow\downarrow}}{2}I$ and $\Bar{\alpha}=\frac{\alpha^{\uparrow}+\alpha^{\downarrow}}{2}\propto S$. Fig. \ref{fig:longitude}(a) shows the relation between the average of non-local voltage and current and by fitting to our V-I model, we can obtain $\Bar{\alpha}\approx 0.032mV/mA^{2}, \Bar{R_{nl}\approx 0.1\Omega},\lambda\approx1.8$. By adjusting the gate voltage, the trending of $\Bar{\alpha}$ is shown in Fig. \ref{fig:longitude}(b) and fitting to the Mott model Seebeck coefficient ($S_{Mott} =k\frac{\partial lnR}{\partial E}$), where k is a constant fitting parameter. 

\begin{figure}

 \includegraphics[width=\columnwidth]{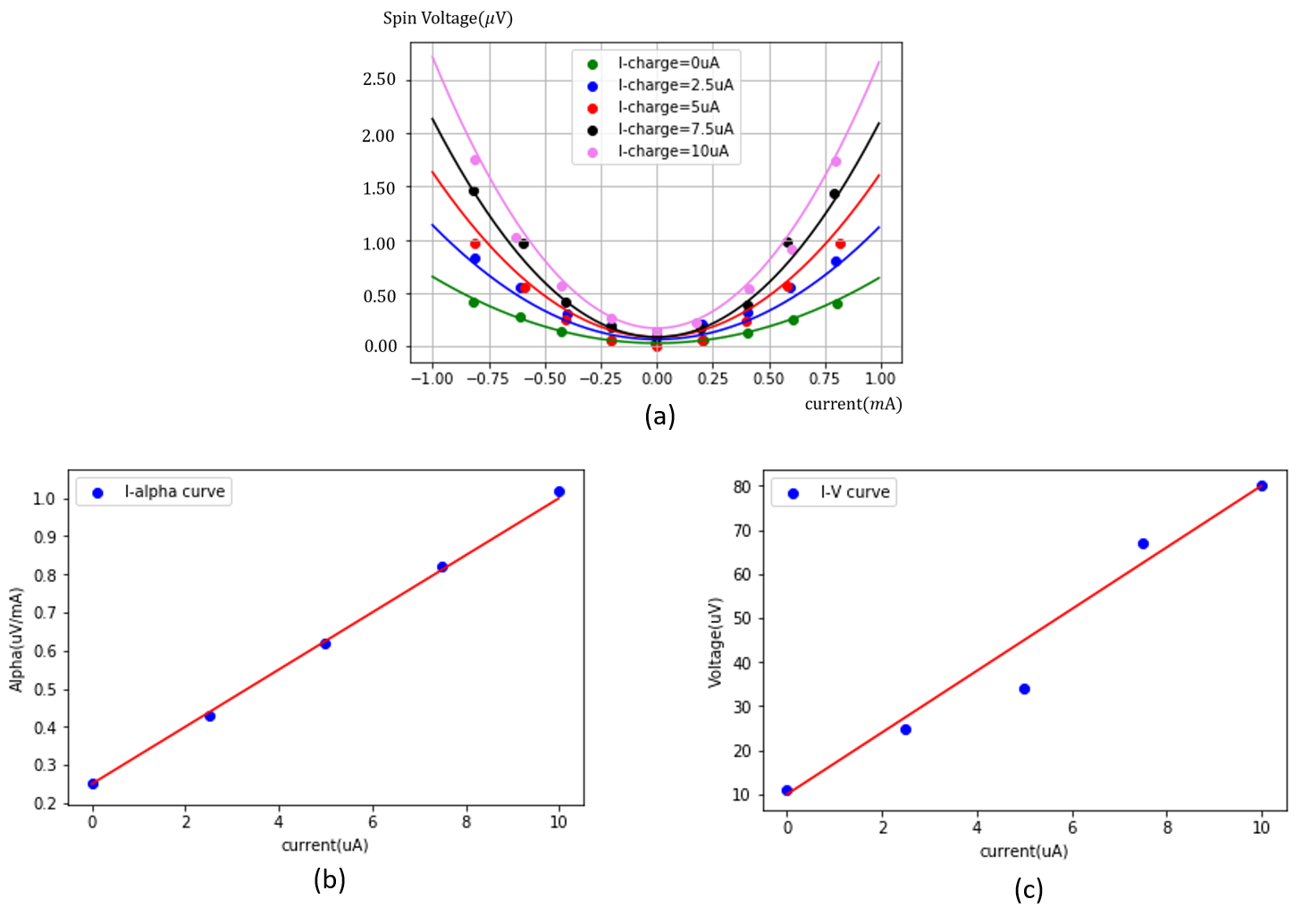}
 \caption{(a) V-I curve when both applying transverse and longitudinal current at T=77K (b) V-$\alpha$ curve when we fix the charge current (c) V-I curve when we fix the heat current }
 \label{fig:combination}
\end{figure}

By combining the transverse and longitudinal current, it is observed that the charge spin signals can be enhanced by thermal spin injection, as shown in Fig. \ref{fig:combination}(a). The transverse heat current we apply along contact 2 is up to 1mA while the longitudinal current we apply is around $10\mu A$. The biases for different charge currents are set to zero for convenience. As we can see in Fig. \ref{fig:combination}(a), the relation between spin voltage and heat current remains quadratic for different charge currents. We assume it is due to the independent sources from thermal and charge channel as:
\begin{gather}
     \delta\mu^{th} = -e\frac{\partial(S\Delta T)}{\partial n}\Delta n \propto -e\frac{\partial(S\Delta T)}{\partial n}R_{nl}I\\
     \delta\mu^{c} = R_{nl}I
\end{gather}
The linear relationship between thermal spin voltage together with charge spin voltage and current, which is predicted by the spin Seebeck effect, is also observed in our experiment, as shown in Fig. \ref{fig:combination}(b)(c). When we increase the applied longitudinal current, both the alpha coefficient and the charge spin voltage increase linearly, as expected. It indicates that the heat and charge current controls the spin voltage coherently.

\section{Conclusion}
In summary, by applying both a transverse and longitudinal current, we investigate the interaction of thermal and charge spin signals on a spin valve device in graphene. We first tested the non-local spin voltage without the heat current and obtained a non-local resistance around 4$\Omega$ and polarization around 18$\%$ according to the Hanle precession measurement, which is consistent with other experimental results in the literature. The thermal spin voltage, which is obtained from transverse thermal spin injection and enhances the charge spin voltage by heat current, can be controlled by both charge spin current and gate voltage. It reaches a maximum close to the Dirac points by adjusting the gate voltage which makes it a promising material in spin caloritronics. The trending of Seebeck coefficient with respect to carrier density can also be observed from longitudinal thermal spin injection and compared with the Mott model. By combining both transverse and longitudinal current together, a linear relationship between non-local spin voltage and spin current is obtained. With the low intrinsic spin-orbit coupling and quadratic enhancement to the spin signals, we expect graphene to play an important role in thermoelectric spin devices in the future. The experimental results demonstrate a new way to control spin transport in SLG by charge, thermal and gate voltage. 

\appendix

\section{Supplementary data: Graphene spin valve device fabrication}
There are various methods in order to fabricate single layer graphene(SLG) which can be categorized into top-down and bottom-up methods \cite{fabrication}. The top-down methods include the 'scotch tape' method \cite{tape}, liquid phase exfoliation \cite{liquid}, laser radiation \cite{laser} and reduced graphite oxide (rGO) method \cite{rgo}. The bottom-up methods mainly focus on epitaxial growth on SiC \cite{sic} and chemical vapor deposition (CVD) on different substrates \cite{cvd}. Here we employ the traditional mechanical exfoliation method (also known as 'scotch tape' method ), which was first discovered by Geim and Novoselov, to fabricate SLG since it imports less impurities and can efficiently produce high quality samples. We first transfer the graphene from  ZYA grade HOPG onto the scotch tape, then pressing the tape on another new adhesive tape to decrease the number of graphite layers until it starts getting transparent under light. At last, a few layers of graphite are transferred onto $SiO_{2}/Si$ substrate with markers and cleaned with acetone, Isopropyl alcohol (IPA) and Deionised (DI) water in order to remove the residues and process for identification.     

To distinguish SLG from other multiple layers of graphite, different technologies can be applied, including optical microscopy identification(expert or deep learning automated method), Raman spectroscopy \cite{raman}, atomic force microscope \cite{tape}. In this experiment, the transferred graphene is first observed under a NiKon optical microscope. We can distinguish the SLG from others by ourselves or a deep learning image segmentation neural network method due to the different contrast and color under appropriate thickness of $SiO_{2}$.  The image of graphene flake under optical microscope can be seen from Fig.\ref{fig:raman}(a). To further investigate and determine the number of graphene layers, Raman spectroscopy \cite{raman} or deep learning based identification by optical microscopy is used for further identification the number of layers of graphene sheet. 

% To further investigate and determine the number of graphene layers, Raman spectroscopy \cite{raman} is used as an integral part of graphene study. This technology provides a reliable method to identify the SLG and defects such as vacancies and doped atoms embedded in graphene. Raman spectroscopy is based on inelastic scattering of photons known as Raman scattering. The interaction of laser light and molecular vibrations generates a so-called Raman shift $\triangle\nu = (\frac{1}{\lambda_{1}}-\frac{1}{\lambda_{2}})$. This shift of energy will give the internal vibration information in the sample. As for graphene, the two most apparent features are the G peak at $\approx 1580cm^{-1}$ and the 2D peak at $\approx 2700cm^{-1}$ as shown in Fig.\ref{fig:raman}(b). The G peak is caused by doubly degenerated zone center $E_{2g}$ while the 2D peak is due to the second order of zone-boundary photons. On the other hand, there is a D peak at $\approx 1350cm^{-1}$ which indicates the number of defects. The SLG can mostly be identified by studying 2D peak in the Raman spectroscopy image. The single component 2D band and low width at half maximum (FWHM) as shown in Fig.\ref{fig:raman}(c) is a clear signature of SLG. In bilayer or multiple layer graphene, the 2D peak contains four or more components and therefore broaden the FWHM during the double resonant(DR) process \cite{raman}. 

\begin{figure}

\begin{subfigure}{.5\textwidth}
  \centering
  % include first image
  \includegraphics[width=\linewidth]{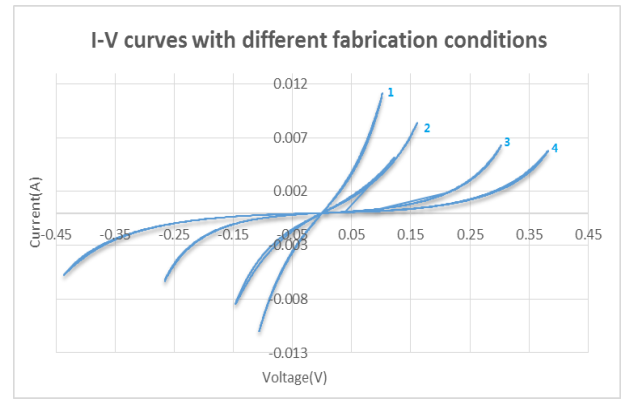}  
  \caption{}
\end{subfigure}
\begin{subfigure}{.5\textwidth}
  \centering
  % include second image
  \includegraphics[width=\linewidth]{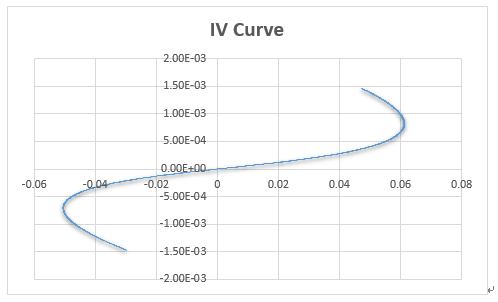}  
  \caption{}
\end{subfigure}%
\newline
\begin{subfigure}{0.5\textwidth}
  \centering
  % include second image
  \includegraphics[width=\linewidth]{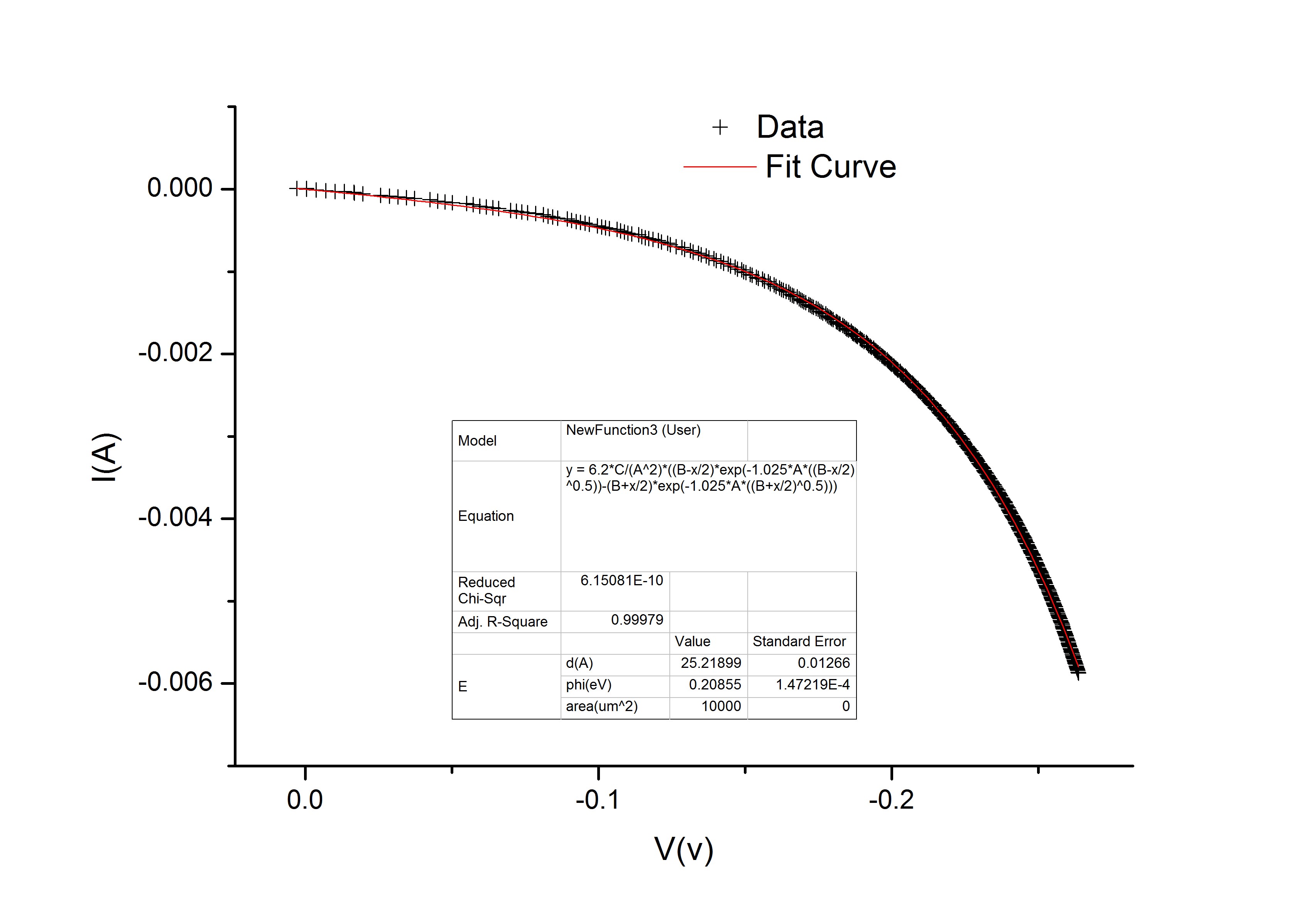}  
  \caption{}
\end{subfigure}%
\begin{subfigure}{0.5\textwidth}%
%   \centering
  % include second image
  \includegraphics[width=\linewidth]{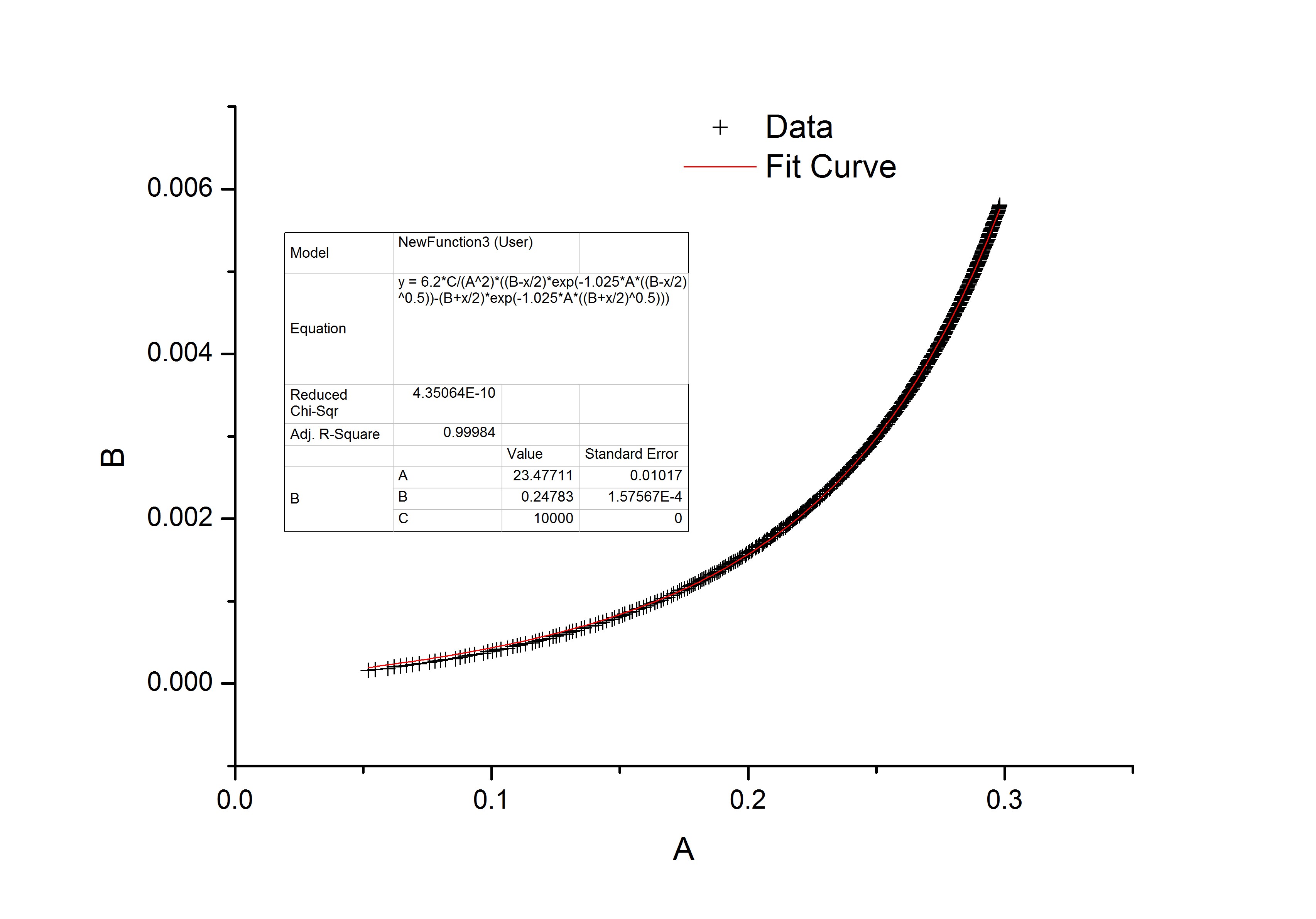}  
  \caption{}
\end{subfigure}%

\caption{(a) 1: 40s oxidation time. 2: 50s oxidation time. 3: Sample
rotation during Al deposition (50s oxidation time). 4:
Sample rotation during both Al deposition and oxidation. (b) Negative resistance due to pinholes when rotation 10rpm during Al deposition.  (c) Fit to Simon model on bottom of curve for best sample. (d) Fit to Simon model on top of curve for best sample. }
\label{fig:deposition}

\end{figure}

The spin valve device for thermal spin injection is described in Fig.\ref{fig:config}. Once the SLG has been identified by optical microscopy and Raman spectroscopy, a 1nm ultrathin tunnel junction \cite{1nm} has to be deposited on the surface. To test the condition for few pinhole magnetic tunnel junction, nine devices were fabricated by DC magnetron sputtering and each device contains four samples. We found nonlinear I-V curves, an indication for an intact tunnel junction. The details strongly depend on the cross sectional area of the contact and the oxidation time. The early devices, of which the cross sectional areas are about 105$\mu m^{2}$, showed a linear I-V, due to high conductivity, as the bias voltage could not reach the nonlinear region. The third device, with a cross sectional area of 104$\mu m^{2}$ and a shorter oxidation time (20s), had a similar linear I-V. By increasing both the cross sectional area and the oxidation time, clear nonlinear I-V curves were observed. However, the devices showed some degree of inhomogeneity. A negative differential resistivity and even a negative resistivity could be observed when high bias voltages were applied to some samples. The main reason for this unusual phenomenon is the low resistivity and low barrier height due to a large amount of pinholes, which causes negative resistance \cite{negative}in crossed-wire devices. To avoid the inhomogeneity and the side contact problem, we rotated the devices during both the Al deposition and the oxygen plasma and sputtered the Al at a 60 degree angle.  By doing this, we could finally make homogeneous ultrathin 1 nm aluminum oxide tunnel junctions, the I-V curve of which can be precisely fitted to the Simon model \cite{simon}. In the Fig.\ref{fig:deposition}, we show the nonlinear I-V curves of several intact tunnel junction samples. The conductance comparison between the samples follows what would be expected from changes in fabrication conditions, as indicated in the legend. 

After the test experiment, we optimized the procedure to deposit ultra-thin $Al_{2}O_{3}$ tunnel barrier with few pinholes. The aluminum oxidation is proceeded by sputter-deposition of a 1 nm thick aluminum layer in 0.19 Pa of Argon at a rotation speed of 10 rpm and post-deposition oxidized in $O_{2}$ plasma to form an aluminum oxide seed layer. Aluminum oxide is then deposited by reactive sputtering from an aluminum target in $Ar/O_{2}$ mixture in an ultra-high vacuum system with rotation speed of 5rpm. An oxygen plasma is generated from oxygen gas using a DC-powered ion source. The deposition parameters are optimized to give an aluminum oxide layer with the best protective/insulating properties: the deposition power (20W), the deposition rate (0.169$\si{\angstrom}/s$), back pressure ($5.9\times10^{-6}Pa$) oxygen partial pressure (flow rate 10 sccm), sputtering gun power (20W), substrate RF bias power (50W), and rotation speed ($Al:10rpm,O_{2}:5rpm$).

\begin{figure}

\begin{subfigure}{.5\textwidth}
  \centering
  % include first image
  \includegraphics[width=\linewidth]{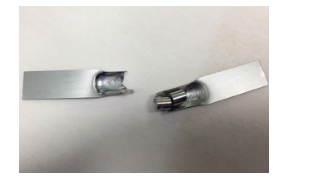}  
  \caption{}
\end{subfigure}
\begin{subfigure}{.5\textwidth}
  \centering
  % include second image
  \includegraphics[width=\linewidth,height=3cm]{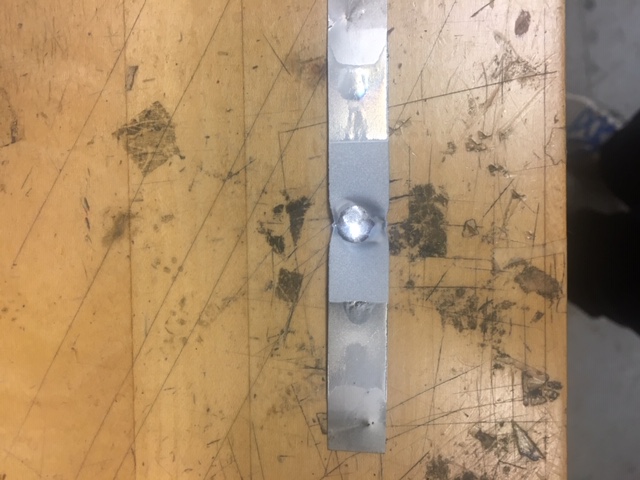}  
  \caption{}
\end{subfigure}
\begin{subfigure}{0.5\textwidth}
  \centering
  % include second image
  \includegraphics[width=\linewidth]{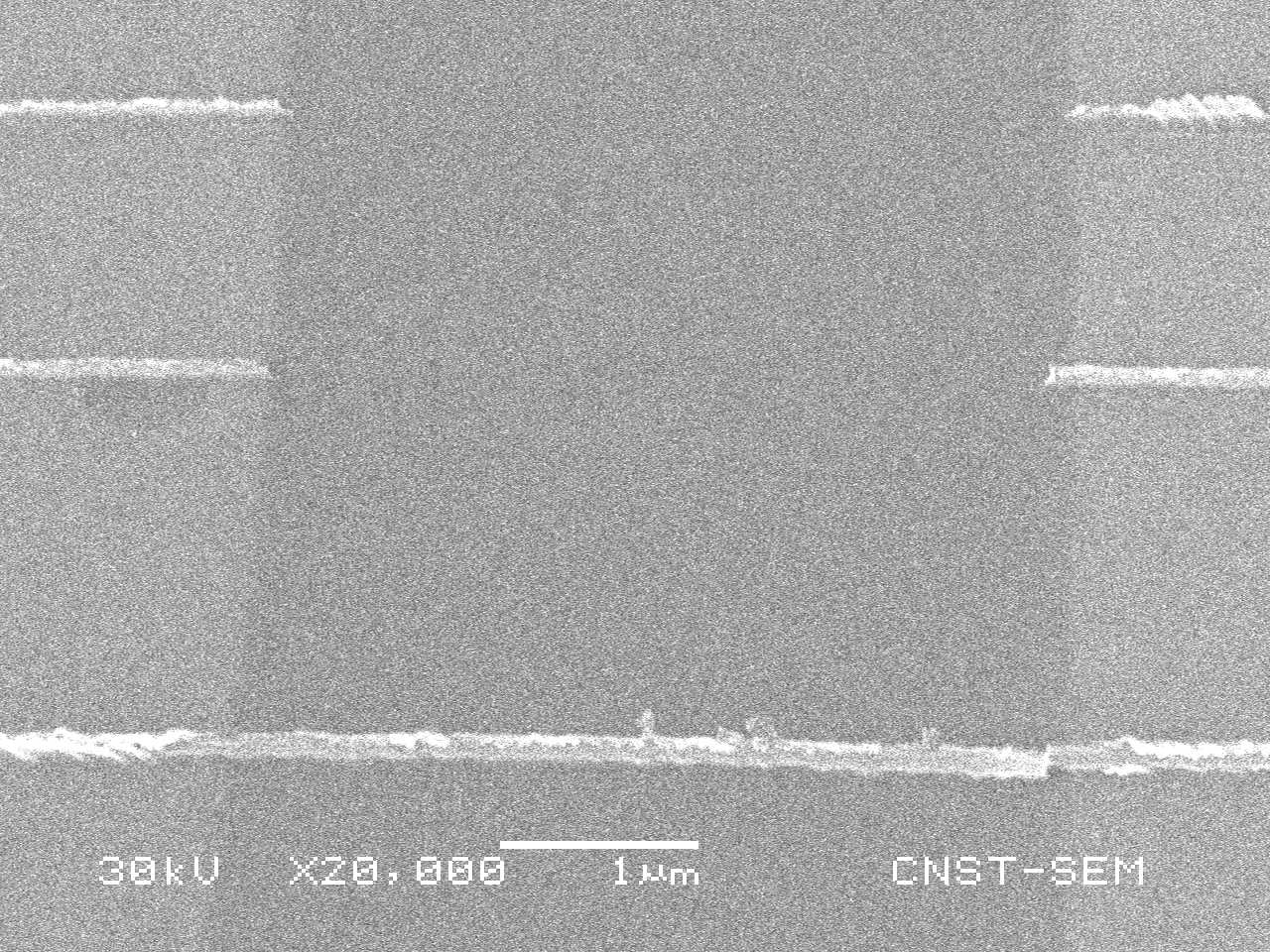}  
  \caption{}
\end{subfigure}
\begin{subfigure}{0.5\textwidth}%
  \centering
  % include second image
  \includegraphics[width=\linewidth]{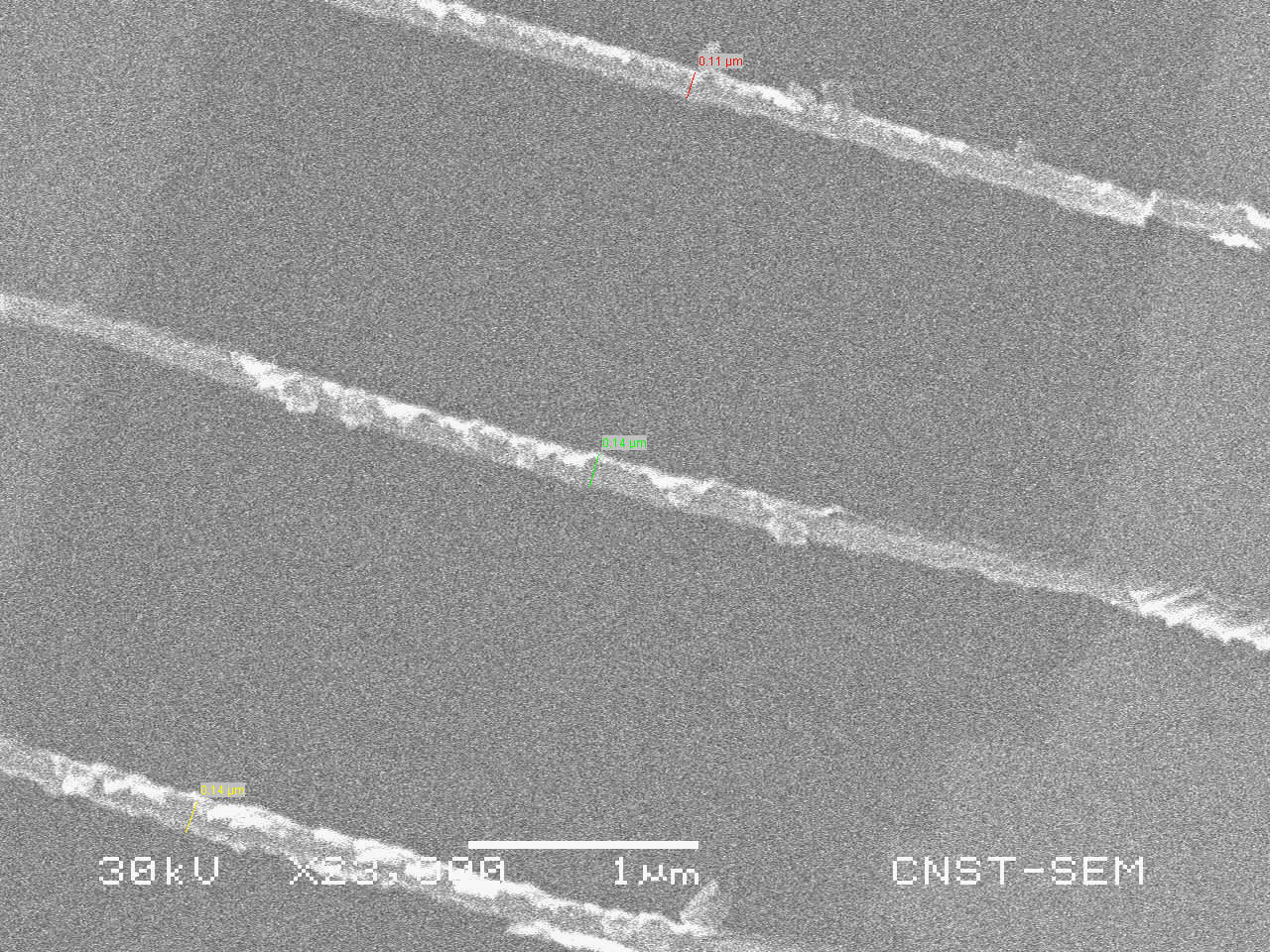}  
  \caption{}
\end{subfigure}
\caption{(a) Tungsten boat breaking down due to cobalt alloy. (b) Tungsten boat with alumina coated.  (c) Cobalt does not adhere to graphene surface due to unstable deposition rate.  (d) Cobalt successfully adhere to surface due to stable deposition rate.  }
\label{fig:cobalt}
\end{figure}

In the last step of device fabrication, Cobalt is deposited on top of the magnetic tunnel barrier by thermal evaporation. However, the high melting temperature and the tendency to alloy with refractory metal made the deposition difficult. While the tungsten boat is heated and cobalt is beginning to melt, it is easy to alloy with the boat and results in boat breaking at last as show in Fig.8. To prevent alloying with the tungsten boat, we use an alumina-coated tungsten boat to deposit the cobalt film. On the other hand, we have to carefully control the current for maintaining a constant deposition rate at about $6\si{\angstrom}/s$. Otherwise, the cobalt will not adhere to the surface tightly as show in Fig.\ref{fig:cobalt}.

In the summary, we made two good samples out of about two hundred broken samples following these three fabrication steps including blind electron lithography, magnetron sputtering and cobalt deposition. The sample we made is shown in Fig.\ref{fig:config}(e)

\end{document}